# Decoupled Moderators – Do we always need them?
# or: A New Approach for Pulse Shaping

A.A. Parizzi [1,2*], W.-T. Lee [1*], G.P. Felcher [3], F. Klose [1*†]

[1] Spallation Neutron Source, Oak Ridge National Laboratory, USA

[2] Electrical Engineering Department, Federal University of Rio Grande do Sul, Brazil

[3] Materials Science Division, Argonne National Laboratory, USA

[*] Current address: Intense Pulsed Neutron Source, Argonne National Laboratory, USA

[†] Corresponding author's email: FKlose@anl.gov

## Abstract

At a spallation neutron source, the wavelength of a neutron is generally determined by its time-of-flight (TOF) from the source to the detector. The achievable precision is ultimately limited by the non-zero emission time-width of the source/moderator system. Particularly, coupled moderators that are used to produce high intensity cold neutron beams show long "tails" in the intensity/time distribution for all wavelengths. For this reason much less intense decoupled moderators are often used for instruments that require high resolution.

We present here a novel technique for dynamic energy filtering of thermal and cold polarized neutron beams. The device consists of a polarizer/analyzer system and an energy-selective spin-resonance flipper. By drifting its resonance condition in synchronization with the TOF, the filter controls the energy resolution of a broad bandwidth TOF neutron pulse. Calculations show that substantial resolution gains are obtainable by applying this technique to a TOF neutron reflectometer.

## 1. Introduction

One of the most difficult problems in neutron scattering is the optimal selection of wavelength resolution for each experiment. Given a constant wavelength (CW) instrument, resolution is conventionally determined by the choice of a monochromating crystal. For time-of-flight (TOF) instruments at reactor sources, the wavelength-resolution can be tuned by changing the speed of a chopper and/or the width of the chopper window. For TOF instruments at spallation neutron sources, the wavelength resolution is determined by a combination of the moderator architecture and the instrument length. Many instruments at pulsed neutron sources that utilize broad bandwidth would gain in versatility if the wavelength-resolution could be tuned by a secondary device, allowing the user to trade intensity for resolution and vice versa.

In short-pulse spallation sources [1], neutrons are created by the impact of a proton beam onto a target material within a very small time interval ($\Delta t < 1\mu s$). These high-energy neutrons are slowed down by a moderator to the energies suitable for use in neutron scattering experiments. In order to gain higher intensity, the target is surrounded by a reflector assembly, which backscatters into the moderator neutrons that would otherwise be lost. There are two classes of moderator-reflector arrangements: "decoupled", in which an absorbing sheet prevents late-returning slow neutrons from entering the moderator, and "coupled", in which all neutrons, including those that thermalize and live longer in the reflector, can enter the moderator. By eliminating the absorber, coupled moderators provide several times higher neutron fluxes than decoupled moderators. Coupled moderators, however, achieve the higher flux at the expense of causing a very long "tail" in the intensity/time distribution of each wavelength. This tail degrades the resolution of the experiments, but it can be removed, if necessary, by means of a wavelength-resolution tuning device, the TOF spin-resonance energy filter, which is proposed below.

An instrument that would certainly benefit from large wavelength resolution flexibility is a neutron reflectometer. Here, resolution requirements can strongly depend on the sample under investigation. Typically, monolayer films have low-resolution requirements (10 - 15%) while a high-resolution capability (1% or even better) is desirable for measurements on films of several thousand-angstrom-thickness. If one chooses a high-resolution monochromator for a CW instrument, one gives up most of the potential intensity for low-resolution measurements. The same applies for a TOF reflectometer at a pulsed source. While low resolution measurements can benefit from the high flux provided by a coupled moderator, high resolution has typically been achieved by placing the instrument on a decoupled moderator where the effective neutron beam is intrinsically less intense or by lengthening the flight-path at the expense of the bandwidth. Using instead a secondary energy filter to tune the energy resolution allows the use of reflectometers at high-intensity coupled moderators for both high-resolution and low-resolution measurements.

## 2. The wavelength resolution at pulsed neutron sources

In order to demonstrate the capabilities of the energy filter, we have calculated the potential resolution gain for a reflectometer being designed for the Spallation Neutron Source project. The SNS *Magnetism Reflectometer* [2] will view a coupled moderator of liquid hydrogen at a nominal temperature of 20 K. The performance of the moderator has been calculated numerically for a set of wavelengths in the range from 0.029 Å to 90.447 Å, by using the MCMP code [3]. We have fitted the performance analytically in the wavelength range that is relevant for reflectometry (0.7 Å to 14 Å) by means of a modified Ikeda-Carpenter model [4].

Fig. 1 shows the neutron beam intensity at the face of the moderator as a function of emission time and wavelength. The moderator emits neutrons for several hundred microseconds after the peak of the pulse emission time distribution. The inserts show the time dependence of the intensity for 3.12 Å and 9.35 Å neutrons. One characteristic of the moderator is that the peak emission time is progressively delayed for longer wavelengths, relative to the time $t = 0$ when the proton pulse strikes the target (the wavelength dependence of the time of the peak intensity is plotted as a dashed line). A second characteristic of the moderator emission is an increase of its full-width-half-maximum intensity with wavelength, which is approximately linear within the studied range. These two features are not the main limiting factors for the wavelength-resolution: a more serious problem is the presence of a long "tail" at later emission times. At the detector, neutrons of shorter wavelengths from the “tails” are not distinguished from longer wavelengths neutrons.

Fig. 2 shows an intensity contour plot as a function of time-of-flight and wavelength seen by a detector at a distance of 19 m from the moderator - the approximate distance estimated in our reflectometer design. In order to demonstrate the effect as clearly as possible, the intensities at each arrival time have been normalized to unity at their peak value. The inserts show two cuts, respectively for a time-of-flight of 15 ms and 45 ms. For 15 ms the maximum intensity occurs at a wavelength of 3.12 Å and for 45 ms the maximum is at 9.35 Å. For both cases, there is a significant contribution of neutrons with shorter wavelengths. The contour lines in Fig. 2 are practically parallel (except for wavelengths shorter than 4 Å). This means that the "tail" neutrons contribute to the broadening of the wavelength-distribution by essentially the same absolute amount for each time-of-flight channel. Long-wavelength neutrons contain a relatively smaller wavelength distribution in their corresponding time channels. This is why better wavelength resolution can be achieved with long-wavelength neutrons. Both inserts in Fig. 2 display the wavelength axis over a 5% range relative to the peak wavelength.

## 3. The spin-resonance energy filter

In 1962 G.M. Drabkin proposed [5], and experimentally confirmed in 1968 [6] a new way for selecting the velocity of polarized neutrons. The basic idea was to guide a polarized neutron beam (either the direct moderator spectrum or a spectrum already pre-monochromated by a mosaic crystal) through an electromagnetic resonator. This resonator flips the neutron spin of a narrow wavelength band within the incoming wide spectrum. The flipped fraction of the beam is separated from the rest by an appropriate polarization analyzer. Effectively, the device acts as a wavelength band-pass filter.

Fig. 3 shows a possible implementation of the spin-resonance energy filter (for a thorough discussion of similar devices see Ref. [7]): A polarizing supermirror reflects spin-up neutrons into the resonator. At the exit of the resonator, a transmission-type spin-analyzer filters (reflects away) the spin-up neutrons, and passes only those neutrons that have been flipped by the resonator. The resonator itself consists of a zigzag folded, DC current-carrying foil, that creates magnetic fields $\pm H_{per}$, perpendicular to the flight path $z$ of the neutrons and alternating their direction along $z$. In addition, a uniform magnetic guide field $H_0$ is present, either perpendicular to both $H_{per}$ and the neutron flight path, or along the neutron flight path. The magnitude of $H_0$ determines the angular precession frequency of the neutron magnetic moment around the $H_0$ axis. This Larmor precession frequency is $\omega_o = \gamma_n \cdot H_o$ ($\gamma_n$ = 1.83 x $10^4$ rad s$^{-1}$ Oe$^{-1}$ is the gyromagnetic ratio of the neutron).

When a neutron of velocity $v$ passes through the resonator, the alternating field $H_{per}$ acts with a frequency $\omega = \pi \cdot v / a$ on the neutron spin, where $a$ is the distance between the current sheets, neglecting the foil thickness. If $H_{per}$ is such that:

$$H_{per} = \frac{v_{res} \cdot \pi}{L \cdot \gamma_n} \tag{1}$$

where $L$ is the length of the resonator, the spin-flip probability approaches 100 % for those neutrons having a "resonance" velocity $v_{res}$ that matches the Larmor frequency:

$$\omega_{res} = \frac{\pi \cdot v_{res}}{a} = \omega_o = \gamma_n \cdot H_o \qquad (2).$$

If the constraints described above are kept for both fields, then the wavelength dependence of the flipping probability is approximately described by the following equation:

$$P(\lambda) = \frac{\left(\frac{1}{M}\right)^2}{\left(1-\frac{\lambda_{res}}{\lambda}\right)^2 + \left(\frac{1}{M}\right)^2} \cdot \sin^2\left\{\frac{M\cdot\pi\cdot\lambda}{2}\sqrt{\left(\frac{1}{\lambda_{res}}-\frac{1}{\lambda}\right)^2+\left(\frac{1}{\lambda_{res}\cdot M}\right)^2}\right\} \text{ [4]} \qquad (3),$$

where $M = L/a$ is the total number of current sheets of the resonator, and

$$\lambda_{res} = \pi \cdot /\, a \cdot m_n \cdot \gamma_n \cdot H_o = \pi \cdot /\, L \cdot m_n \cdot \gamma_n \cdot H_{per} \qquad (4)$$

($m_n$ is the mass of the neutron). Note that, in this case, the constraint between the fields is determined by the relation $H_o / H_{per} = L/a = M$ .

The spin-flip probability function has two characteristic features (see Fig. 4 for a graphical representation). First, the width of the central maximum is inversely proportional to the number of field reversals, which is equivalent to the number of current sheets $M$; i.e., $M$ practically determines the wavelength resolution. Second, there are side maxima with flipping probabilities as high as 12%. The latter can be avoided by several methods, as is shown in Ref. [7].

The energy filter was first considered as a substitute for monochromators at steady state sources [8]. Unfortunately, even with its adjustable resonance wavelength (achieved by parallel changes of $H_0$ and $H_{per}$), flexible selectivity of the wavelength bandpass width (achieved by variation of $M$) and the high flipping-probability capabilities of the resonator, its application at a steady state source demands formidably high efficiency of the supermirror polarizer/analyzer.

Our proposal is to use the filter in a dynamic way, as a "pulse shaper" to tune the wavelength-resolution in TOF experiments at spallation neutron sources. We propose to use the same basic setup (see Fig. 3), but additionally drive the Drabkin resonator fields by fast power supplies to match the time structure of the pulsed source. The calculations, presented in the following, show that limitations of supermirrors with respect to polarization are not an issue for TOF pulse shaping.

### 4. Tailoring moderator pulse shapes with a spin-resonance energy filter

The insertion of the energy filter can be very effective to cut unwanted tails from neutron beams. In TOF instruments, a time-dependent transmission window can be created by drifting the resonance condition with time. This can be achieved by appropriately controlling the magnetic fields $H_o$ and $H_{per}$ inside the resonator (see Fig. 3). Only neutrons that cross the resonator at a certain time and with a specific wavelength are transmitted. Neutrons that do not fulfill these conditions are reflected out by an analyzing supermirror. Note that the rejected neutron stream will not only contain the previously mentioned moderator tail neutrons, but also delayed neutrons from the target and other background neutrons. Therefore, the device may also significantly improve the signal to noise ratio.

To test the effectiveness of the energy filter in removing the unwanted "tail" neutrons, we started by arbitrarily choosing an arrival time of 25365 μs at the detector (which we assume to be positioned at 19m from the moderator). The corresponding peak in the expected wavelength distribution at this time-of-flight is 5.26 Å (see Fig. 5). The initially unpolarized beam is reflected by a supermirror, which preferentially reflects spin up neutrons with a realistic efficiency of 20 to 1 (the corresponding reflectivities are 95.24% for spin up and 4.76% for spin down neutrons). The resulting wavelength distribution at the detector is displayed in Fig. 5a. Next, we insert the magnetic resonator in the polarized beam. The device is placed at 9 m from the moderator and has 300 current sheets separated by 2.2 mm. The corresponding wavelength resolution $\Delta\lambda/\lambda$ is 0.2%. The resonance condition is tuned to $\lambda_{res} = 5.26$ Å via the fields $H_0$ and $H_{per}$. The resulting outgoing spectrum, obtained with the aid of equation (3), is displayed in Fig.5b. Note that, due to the non-symmetric peak shape of the initial wavelength distribution, a secondary maximum appears only at the left side of the spin-flipped spectrum. Finally, we add an analyzer to the setup in order to filter out the narrow bandwidth region around $\lambda_{res}$ that has been flipped. As an analyzer, we have chosen a recently developed white beam neutron spin splitter that allows using the transmitted and reflected beam at the same time [9]. The flipping ratio of the analyzer has been reported to be 25 for the reflected beam (mostly spin-up neutrons) and 100 for the transmitted part of the beam (mostly spin-down neutrons). The spin-resolved transmitted spectra after passing the analyzer are shown in Fig. 5c.

The comparison between the final narrow bandwidth profile and the incoming polarized intensity is displayed in Fig. 6. In both cases, we have added up the contribution of both spin states. One can clearly see that the neutron tail is removed very effectively and that the resulting spectrum is much more symmetric and sharper. Note that the intensity losses in the central part of the initial peak are only minor. An important side effect of the analyzer is that the beam polarization is significantly increased. The flipping ratio after passing the complete energy filter system improves to 51 (corresponding to a beam polarization of 96.1%), compared to a flipping ratio of only 20 that is introduced by the initial polarizer.

To extend the improvement for a particular time-of-flight channel to the entire time-of-flight spectrum, one needs to continuously vary the resonance condition by adjusting the currents which create the resonance fields $H_o$ and $H_{per}$, as $1/t$ between neutron pulses. For the transition from the static (single TOF channel) to the dynamic (all TOF channels) operating mode, there is, however, an important design change necessary that is related to the physical length of the device (typically 0.5m - 1m): In order to achieve a spin flip for a neutron of a particular wavelength, this neutron must encounter an appropriate <u>static</u> combination of $H_0$ and $H_{per}$ all the way through the resonator. On the other hand, there is the requirement to drift the field strengths proportional to $1/t$ in order to match the resonance to the actual peak wavelength. Both conditions can only be fulfilled by introducing field compensation. Calculations show that the guide field in particular requires significant compensation. A solution for this problem is to superimpose an additional guide field $H_0^*$ with linear gradient along the flight path. $H_0^*$ needs to have the same $1/t$ dependence as $H_0$. Fig. 7 shows the resulting spectra with and without using the compensation field $H_0^*$.

To ascertain the effect of beam divergence on the filter performance, we have repeated the calculations above for incident angles within a range of ±1°. This angular range far exceeds the divergence of a neutron beam typically used in measurements. Virtually no angular dependence of the filter performance can be found. The filter effectively decouples the energy resolution and the angular resolution, allowing each to be adjusted independently.

### 5. An example: Resolution improvement for a reflectometer

An example for a potential resolution gain in a neutron scattering experiment is given in Fig. 8. Here, we have simulated a reflectivity measurement from a 7000 Å Ni film on a Si substrate. The energy filter is located at 9 m from the source and *M*=300 for the resonator. The displayed q-range corresponds to a nominal wavelength-range of 6.0 Å - 5.3 Å The angle of incidence was assumed to be 0.57°. In Fig. 8a, we compared the reflectivity curve without resolution effects (thin line) with a convolution of this curve with the original SNS moderator wavelength/time structure (bold line). The insertion of the described energy filter gives rise to the reflectivity presented in Fig. 8b. The resolution is clearly improved to the extent of bringing the experimental curve close to the bare Fresnel reflectivity.

The use of the spin-resonance energy filter is beneficial also from another aspect: Conventionally the experimental reflectivities are assumed to be characterized by a resolution function of the type:

$$\frac{\Delta q}{q} = \text{constant} = \sqrt{\left(\frac{\Delta\theta}{\theta}\right)^2 + \left(\frac{\Delta\lambda}{\lambda}\right)^2} \qquad (5),$$

with both the angular resolution, $\Delta\theta/\theta$, and the wavelength resolution, $\Delta\lambda/\lambda$, representing the relative width of gaussian functions constant with the wavelength and with the angle of incidence. If this were the case, measurements taken over a span of wavelengths at different angles could be seamlessly spliced together. It has been seen that this assumption is only approximately valid [10] particularly at small angles. Even if reflectivity data obtained at different angles were presented separately, their fitting to a model should be accomplished with the help of Rietveldt methods [11], which would use correct resolution functions for angle and wavelength.

Our proposal to utilize the energy filter as a "tail cutting" device goes a long way toward alleviating these problems. Compared to other techniques that have been recently suggested for this function, it has the following general advantages: a) It works very well for instruments that use wide-bandwidth neutron beams. This is not true for a tail cutting chopper system that can only handle a narrow bandwidth region. b) The resolution can be altered rapidly by electronically changing the number of active current sheets. Disk choppers need on the order of minutes for re-phasing, an $E_o$ chopper much longer if slit packages need to be changed. c) It is practically insensitive to the angular divergence of the beam. In contrast to crystal - or multilayer - based monochromators, angular and wavelength resolution are effectively decoupled in the energy filter.

**Acknowledgements**

We are grateful to M.M. Agamalian, C.F. Majkrzak and L. Passell for enlightening discussions about their experiences with Drabkin spin-resonance devices. We appreciate the cooperation of NIST and Brookhaven Nat. Lab. to provide us with prototype resonance-filter devices for future experiments. We also acknowledge fruitful discussions with J.M. Carpenter and E.B. Iverson, who also supplied us with Monte Carlo results of the expected performance of the SNS coupled hydrogen moderator. This work is supported by the US Department of Energy under contract No. DE-AC05-000R22725 (SNS Project). The work of one of us (GPF) was supported by the Department of Energy, Office of Science, under contract 31-109-ENG-38.

## References

[1] J.M. Carpenter, Nucl. Instr. and Meth. **145** , 91 (1977).

[2] F. Klose, Design Criteria Document for the SNS Magnetism Reflectometer, SNS-07050000-DC00001 (2000). (Document is available at: www.sns.anl.gov)

[3] E.B. Iverson, Detailed SNS Neutronics Calculations for Scattering Instruments Design, SNS/TSR – 203, POI5 configuration, 13 September 2000.

[4] S. Ikeda, J.M. Carpenter, Nucl. Instr. and Meth. A **239**, 536 (1985).

[5] G.M. Drabkin, Zh. Eksp. Teor. Fiz. **43**, 1107 (1962).

[6] G.M. Drabkin, V.A. Trunov, V.V. Runov, Sov. Phys. JETP **54** (1968) 194 - English version of Zh. Eksp. Teor. Fiz. **54**, 362 (1968).

[7] M.M. Agamalian, G.M. Drabkin, V.I. Sbitnev, Physics Reports (Review Section of Physics Letters) **168**, 265 (1988).

[8] C.F. Majkrzak, C.J. Glinka, S.K. Satija, in Proceedings of SPIE Vol. 983 Thin Film Neutron Optical Devices (1988), page 129.

[9] T. Krist, F. Klose, G.P. Felcher, Physica B **248**, 372 (1998).

[10] L.J. Norton, E.J. Kramer, F.S. Bates, M.D. Gehlsen, R.A.L. Jones, A. Karim, G.P. Felcher, and R. Kleb, Macromolecules **28**, 8621 (1995).

[11] M.H. Mueller, R.A. Beyerlein, J.D. Jorgensen and T.O. Brun, J. Appl. Cryst. **10**, 79 (1977).

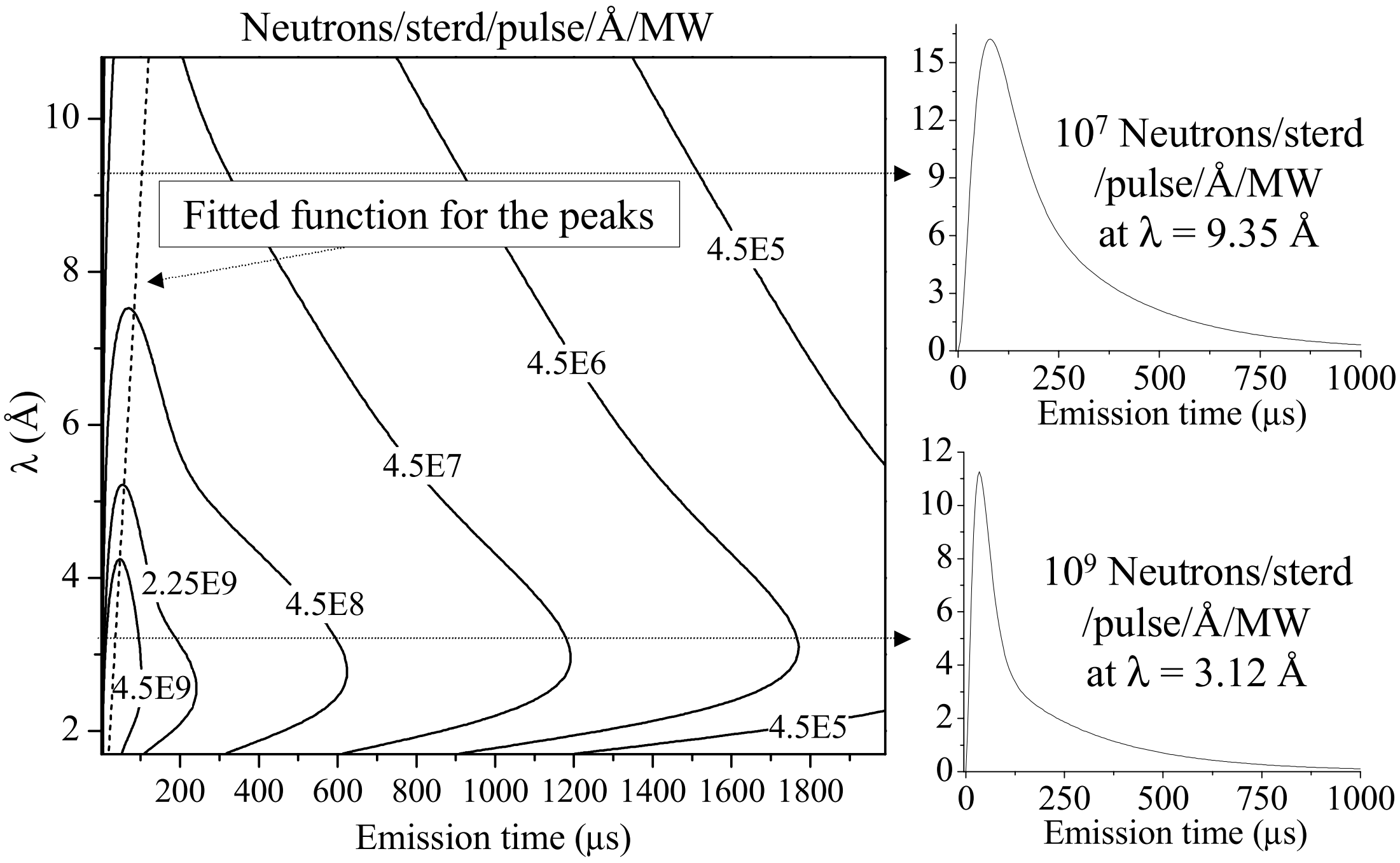

Fig. 1 Calculated time/wavelength-dependence of the neutron flux at the face of the SNS coupled liquid hydrogen moderator. The smaller diagrams show a cut through the contour plot for two particular wavelengths (3.12 Å and 9.35 Å, see arrows). The dashed line indicates the position of peak intensity for each wavelength.

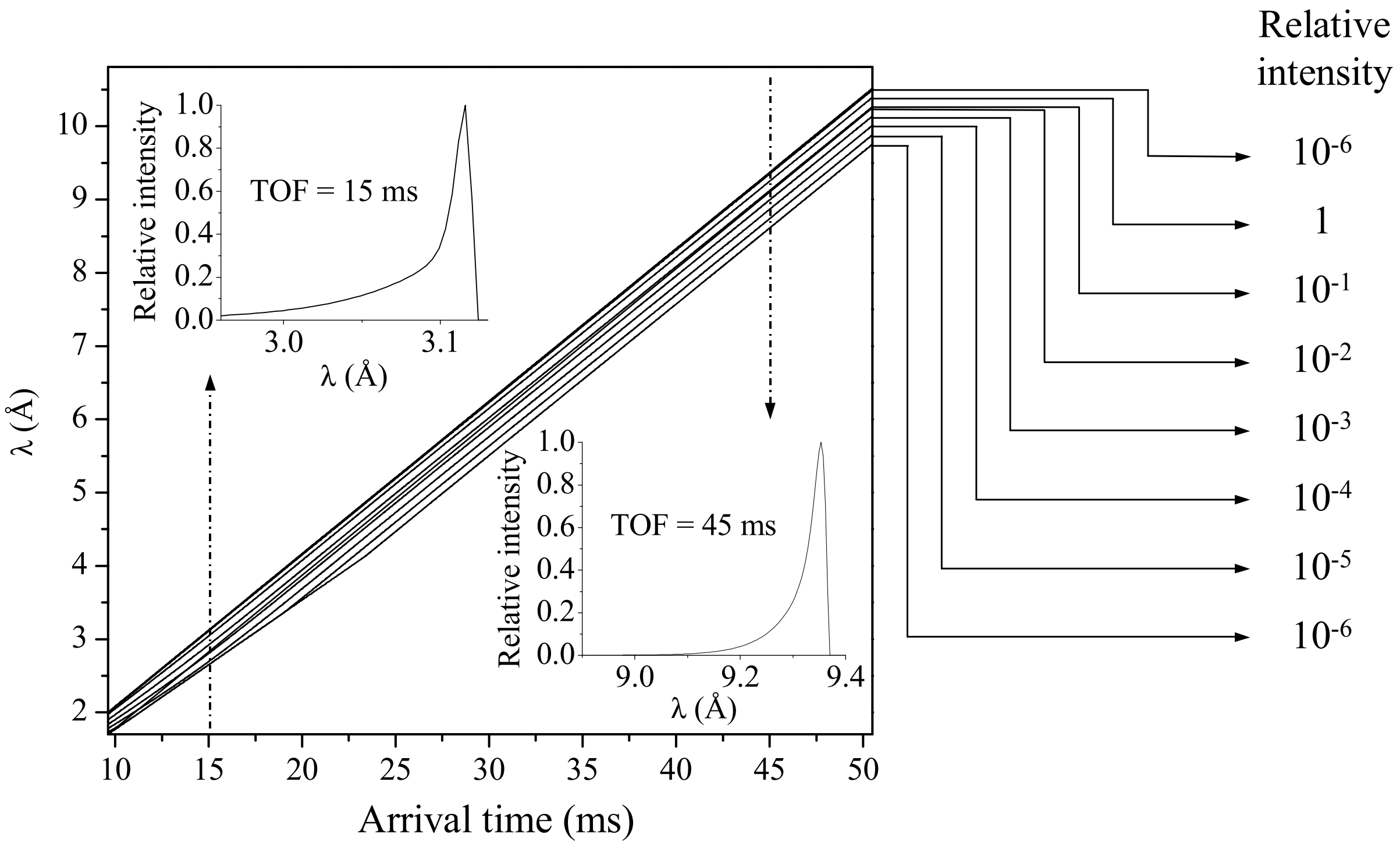

Fig. 2 Calculated time-wavelength dependence of the neutron intensity at the detector position. The data is shown normalized to the peak intensity for each TOF channel (see text). The inserts show cuts through the contour plot at TOF of 15 ms and 45 ms (see dashed arrows).

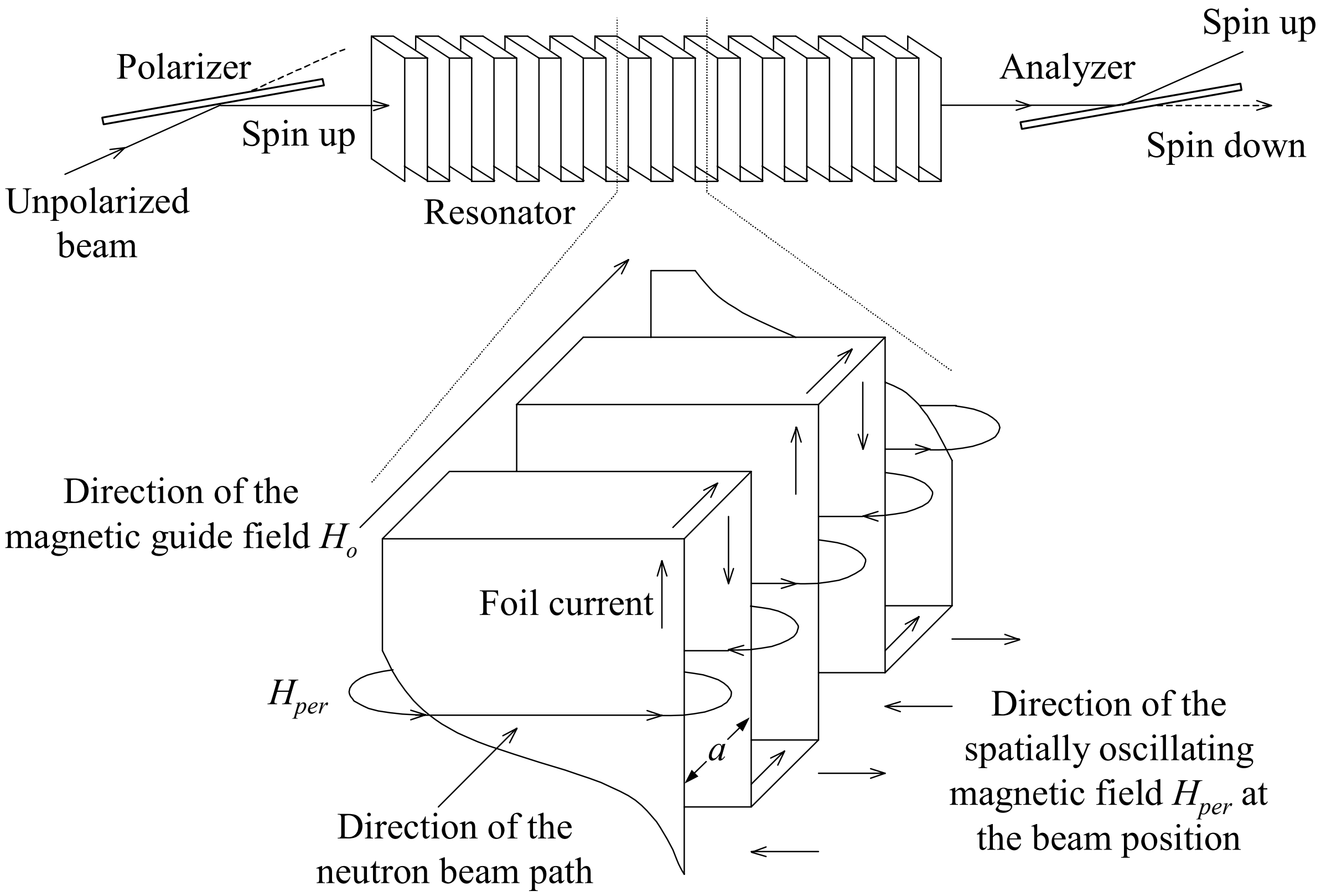

Fig. 3 Spin Resonance Energy Filter and a section of the Drabkin resonator.

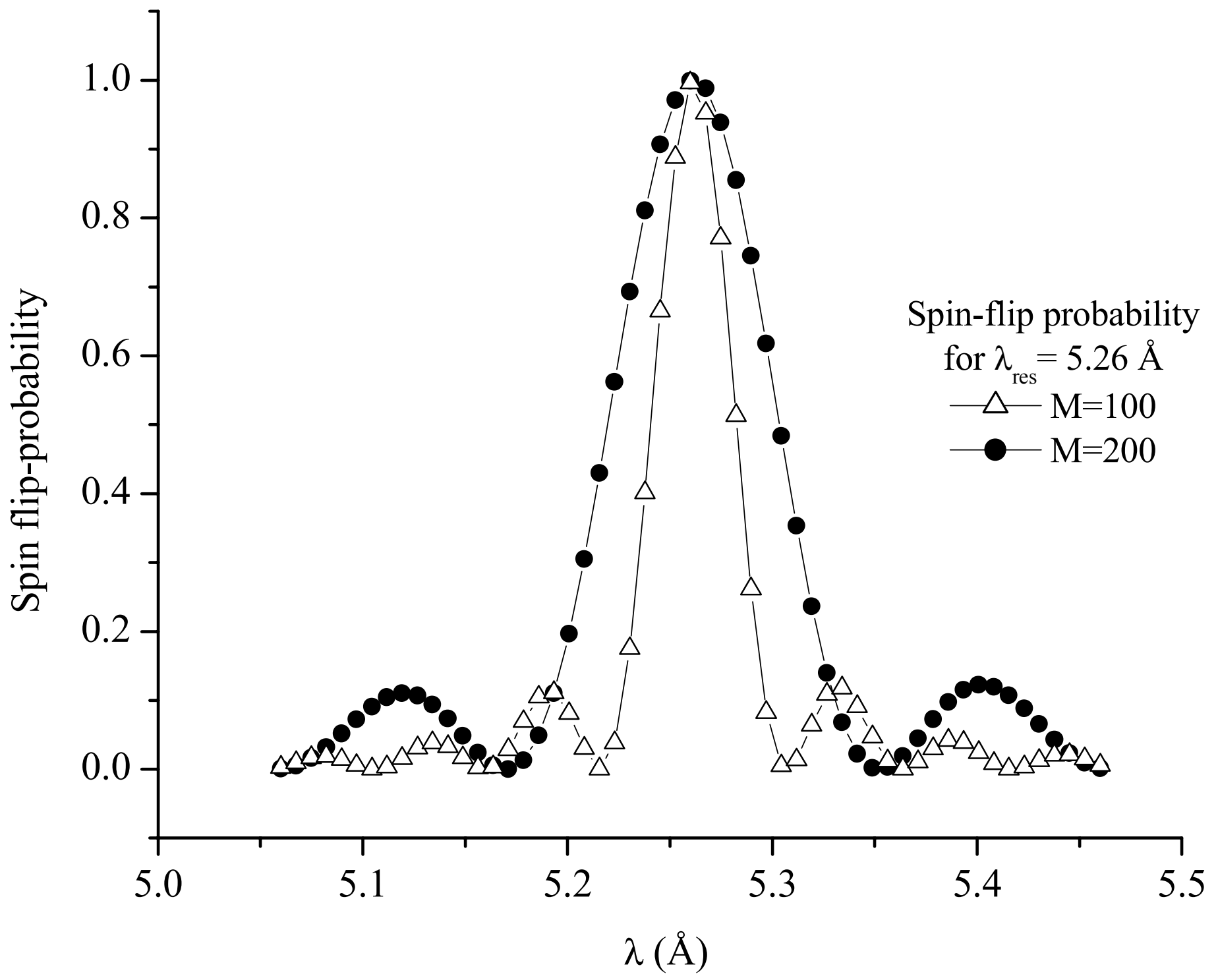

Fig. 4 Wavelength-dependence of the spin-flip probability for different numbers of current sheets. In this example, the resonance wavelength $\lambda_{res}$ is chosen to be 5.26 Å.

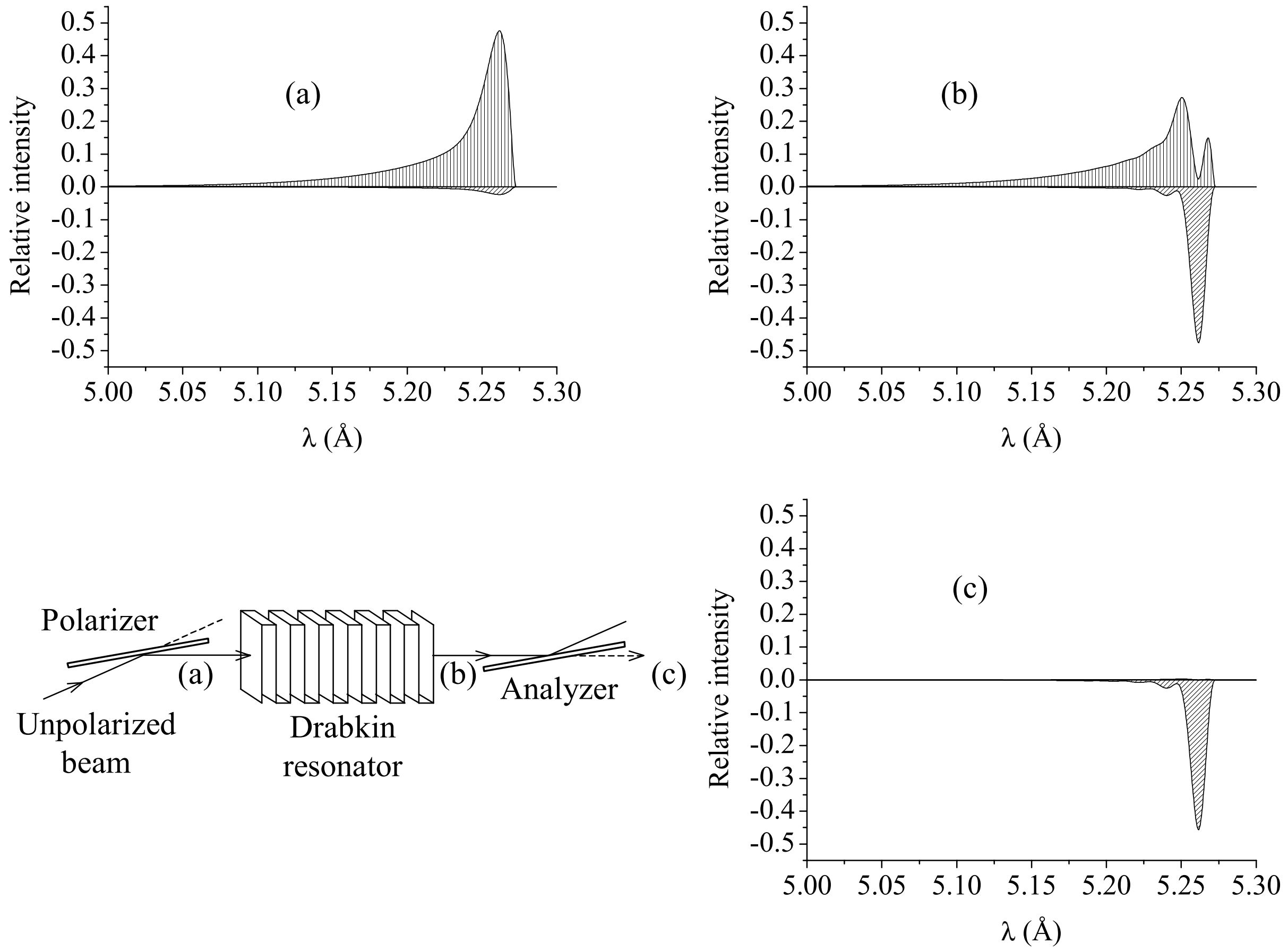

Fig. 5 Spin-resolved relative intensities at the detector for TOF = 25.265 ms. The diagrams show the effect of inserting: (a) – the polarizer. (b) – polarizer and Drabkin resonator. (c) – polarizer, Drabkin resonator, and analyzer.

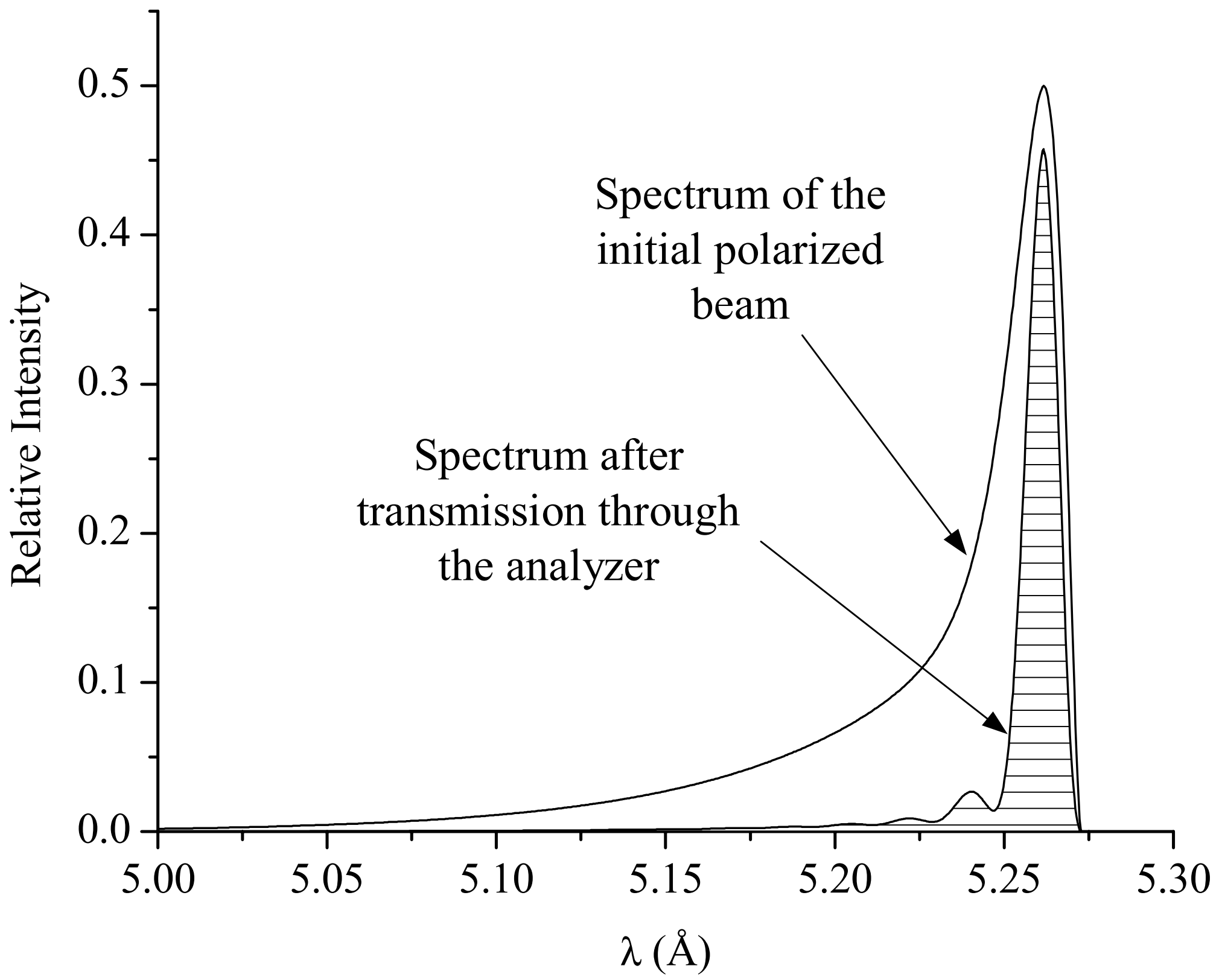

Fig. 6 The resultant conditioned beam profile compared to the initial polarized beam profile.

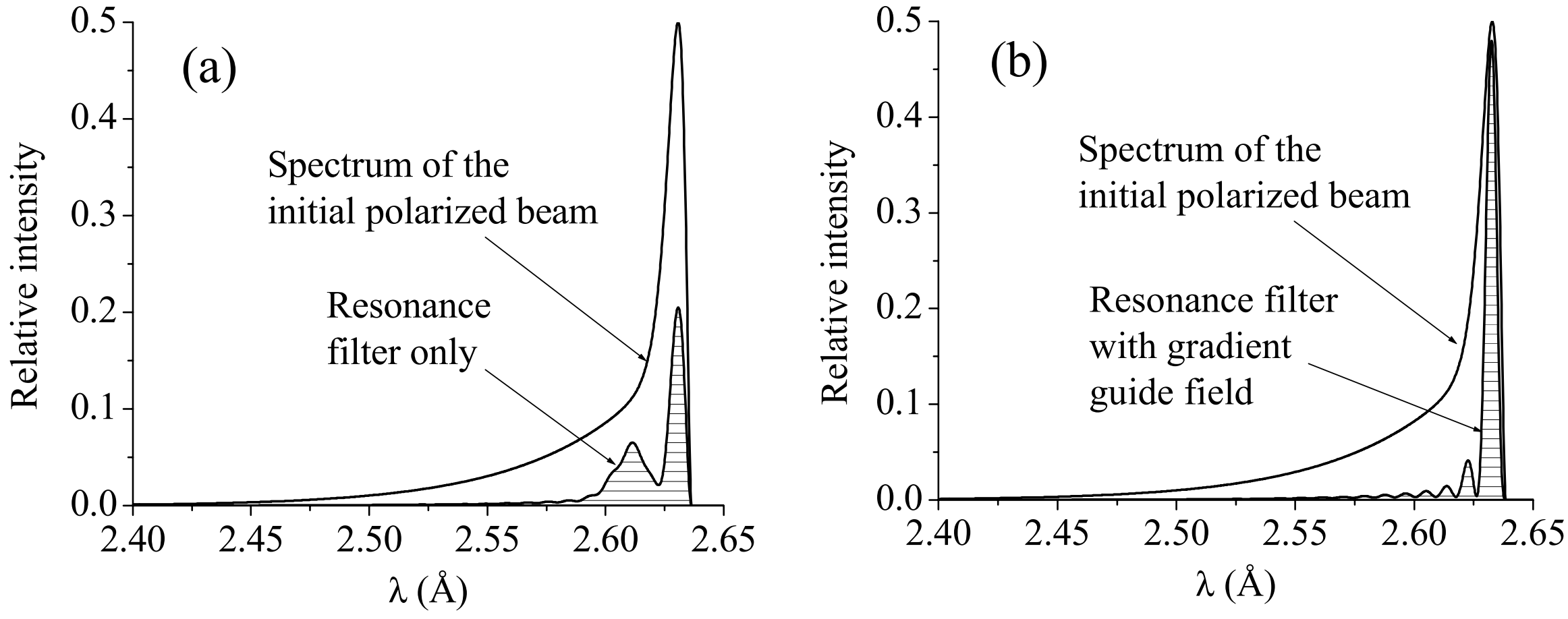

Fig. 7 Spectrum of the beam transmitted through the analyzer: (a) degraded by drifting the resonance without compensation; (b) recovered by imposing a compensating time-varying guide field with linear spatial gradient along the flight path.

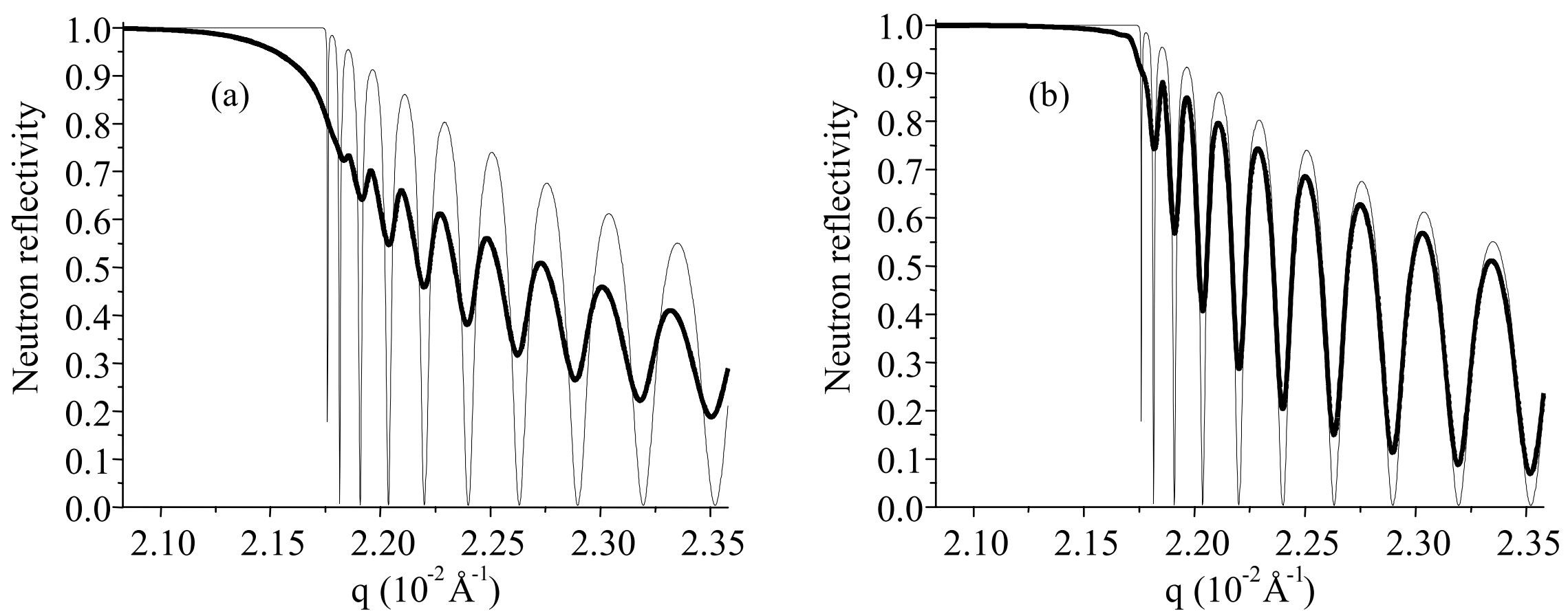

Fig. 8 Simulated reflectivity for a 7000 Å Ni film on a Si substrate, with the parameters as discussed in the text. The thin line in both figures are theoretical reflectivity without resolution effects. The thick lines are (a) with the filter off; (b) with the filter activated.